\documentclass[reprint, amsmath, amssymb, aps,
superscriptaddress, floatfix]{revtex4-2}

\usepackage{color}
\usepackage[dvipsnames]{xcolor}
\usepackage[normalem]{ulem}
\usepackage{float}
\usepackage{graphicx}
\usepackage{dcolumn}
\usepackage{bm}
\usepackage{braket}
\usepackage{amsmath}
\usepackage{algorithm}
\usepackage{algpseudocode}
\DeclareMathOperator*{\argmax}{arg\,max}

\newcommand{\change}[1]{{\color{Black}#1}}
\newcommand{\newstuff}[1]{{\color{Blue}#1}}

\begin{document}

\preprint{APS/123-QED}

\author{Joseph G. Smith}
\email{jgs46@cam.ac.uk}
\affiliation{Cavendish Laboratory, Department of Physics, University of Cambridge, Cambridge, CB3 0HE, United Kingdom}
\affiliation{Hitachi Cambridge Laboratory, J. J. Thomson Ave., Cambridge, CB3 0HE, United Kingdom}
\author{Crispin H. W. Barnes}
\email{chwb101@cam.ac.uk}
\affiliation{Cavendish Laboratory, Department of Physics, University of Cambridge, Cambridge, CB3 0HE, United Kingdom}
\author{David R. M. Arvidsson-Shukur}
\email{drma2@cam.ac.uk}
\affiliation{Hitachi Cambridge Laboratory, J. J. Thomson Ave., Cambridge, CB3 0HE, United Kingdom}

\title{An adaptive Bayesian quantum algorithm for phase estimation}

\begin{abstract}

\change{Quantum-phase-estimation algorithms are critical subroutines in many applications for quantum computers and in quantum-metrology protocols. These algorithms   estimate the unknown strength of a unitary evolution. By using  coherence or entanglement to sample the unitary $N_{\mathrm{tot}}$ times, the variance of the estimates can scale as $O(1/{N^2_{\mathrm{tot}}})$, compared to the best ``classical'' strategy with $O(1/{N_{\mathrm{tot}}})$. The original algorithm for quantum phase estimation cannot be implemented on near-term hardware as it requires large-scale entangled probes and fault-tolerant quantum computing. Therefore, alternative algorithms have been introduced that rely on  coherence and statistical inference. These algorithms produce quantum-boosted phase estimates without inter-probe entanglement. This family of phase-estimation algorithms have, until now, never exhibited the possibility of achieving  optimal scaling $O(1/{N^2_{\mathrm{tot}}})$.  Moreover, previous works have not  considered the effect of noise on these algorithms. Here, we present a coherence-based phase-estimation algorithm which can achieve the optimal quadratic scaling in the mean absolute error and the mean squared error. In the presence of noise, our algorithm  produces errors that approach the theoretical lower bound. The optimality of our algorithm stems from its adaptive nature: Each step is determined, iteratively, using a Bayesian protocol that analyses the results of previous steps.}

\end{abstract}

\maketitle

Quantum phenomena can be used to improve measurement techniques beyond what is achievable with classical techniques using  similar resources. One example of this is quantum phase estimation. Quantum phase estimation is used as: a subroutine of many quantum algorithms \cite{QAE, QAE2, shor, Qcount}; a component in gravitational-wave detection \cite{grav_waves}; a method to measure time \cite{clocks}; and as a tool to compute ground-state energies \cite{VQE, VQE2}. Phase estimation is the estimation of an unknown phase, $\theta$, that a unitary operation $\hat{U}(\theta)$ applies to a quantum state. Traditional techniques evolve separable probes via $\hat{U}(\theta)$ and can lead to an estimate with variance bounded by the standard quantum limit, $O(1/{N_{\mathrm{tot}}})$, where $N_{\mathrm{tot}}$ is the number of unitary applications \cite{Lloyd1} \footnote{{Note that this work concerns asymptotic phase estimation, where $N_{\mathrm{tot}}$ is large; non-asymptotic strategies are discussed in \cite{Wilfred}}}. If multiple probes are entangled instead, estimates can be made with variance bounded by the Heisenberg limit, $O(1/{N^2_{\mathrm{tot}}})$ \cite{Cramer}. However, the most trivial applications of entangled states in phase estimation yield ambiguous estimates due to the multi-valued nature of the inverse functions used \cite{Me}. Precise, unambiguous estimates of $\theta$ require the use of a phase-estimation algorithm. Such algorithms achieve what is known as point identification of $\theta$'s estimate \cite{point_id}.

The goal of phase-estimation algorithms is to make the best estimate of $\theta$ given a certain amount of resources. The best known example is the quantum-phase-estimation algorithm (QPEA) \cite{QPEA, nielsen}. The QPEA requires large-scale fully entangled states, the application of several controlled versions of $\hat{U}(\theta)$ and the ability to implement the inverse Fourier-transform algorithm \cite{FT}. Although the QPEA  attains Heisenberg scaling and point identification, its requirements put stringent limits on which systems or platforms it can be successfully implemented on \cite{noisy_QPEA, noisy_QPEA2}. This stringency has lead to the development of less-cumbersome phase-estimation algorithms based on statistical inference \cite{MLE, Exp_MLE, RG, Burgh}. These algorithms construct distributions of possible candidate values for the estimate of $\theta$ by iteratively sampling multiple circuits. The circuits constitute several applications of $\hat{U}(\theta)$, but do not necessitate inter-probe entanglement. A shortcoming of previous such algorithms is that they fail to obtain  Heisenberg-scaling in any error other than the single-peak variance \cite{Me}.

Another shortcoming of current phase-estimation algorithms is that few have considered the effect of noise present in realistic devices. Current quantum devices belong to the class of noisy intermediate-scale quantum (NISQ) devices and are prone to environmental noise: qubits suffer from preparation and measurement noise and from circuit-induced decoherence. Such noise often degrades the acquired information and reduces the quantum enhancement from a quadratic boost to a constant-factor improvement \cite{noisy_QFI}. 

In this letter, we present an adaptive Bayesian phase-estimation algorithm. This algorithm obtains, in the limit of large $N_{\mathrm{tot}}$, the Heisenberg scaling of errors when noise is absent, and a theoretical lower bound on errors when noise is present. The errors we consider are the mean absolute error and the mean squared error. Our algorithm attains this performance through adaptive iteration, executing a series of circuits with parameters depending on the results of previous measurements. These circuits require one probe at a time and, hence, no inter-probe entanglement \footnote{However, the probes might be entangled states.}. We  compare our Bayesian algorithm with other phase-estimation algorithms. In numerical simulations we show that our algorithm outperforms previous iterative algorithms, as well as the QPEA for small and large values of $N_{\mathrm{tot}}$.

\begin{figure}
    \centering
    \includegraphics[width=6.5cm]{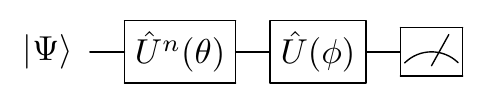}
    \caption{The quantum circuit $(n,\phi)$.}
    \label{fig:circuit}
\end{figure}

\textit{Bayesian phase estimation.---}We consider the unitary operation $\hat{U}(\theta)$ with two eigenstates $\ket{\psi_0}$ and $\ket{\psi_1}$, such that
\begin{equation}
\label{eqn:U}
\begin{split}
    &\hat{U}(\theta)\ket{\psi_0} = \ket{\psi_0} \\
    &\hat{U}(\theta)\ket{\psi_1} = e^{i\theta}\ket{\psi_1}.
    \end{split}
\end{equation}
Here, $\theta \in [0,2\pi]$ is the unknown phase to be estimated. When  measuring $\theta$, it is optimal to do so with probes in an equal superposition of $\ket{\psi_0}$ and $\ket{\psi_1}$ \cite{Lloyd1}:
$\ket{\Psi} = \frac{1}{\sqrt{2}} \left( \ket{\psi_0}  +  \ket{\psi_1} \right)$. Here, we focus on such optimal phase estimation and consider only two-level subspaces of potentially $d$-dimensional unitaries. An estimate of $\theta$ can be achieved through analyzing the output of the quantum circuit in Fig. \ref{fig:circuit}. A single probe is prepared in the state $\ket{\Psi}$ and evolved coherently through $\hat{U}(\theta)$ a number $n$ times. The probe is then evolved a single time through a known phase-shift $\hat{U}(\phi)$ before being measured in the basis $\left\{ \ket{\Psi}, \ket{\Psi^\perp}\right\}$. (Throughout this letter, we  use $(n,\phi)$ to denote such a circuit). This measurement process maximizes the quantum Fisher information (QFI) and leads to asymptotically optimal measurements \cite{Cramer, Lloyd2} 
\footnote{An alternative QFI-maximizing measurement involves a Greenberger-Horne-Zeilinger (GHZ) state \cite{Lloyd1}. However, owing to the extreme noise sensitivity of GHZ states \cite{noisy_entangle}, we only consider coherent unitary applications in this paper.}.
The circuit $(n,\phi)$ is sampled $\nu$ times, and the number of probes in the state $\ket{\Psi}$ is recorded as $x$. This process applies $\hat{U}(\theta)$ a number $n \nu$ times. The probability of $x$ taking a certain value is given by the binomial distribution:
\begin{equation}
\label{eqn:p_x}
    p(x|n,\phi,\nu,\theta) = \begin{pmatrix} \nu \\ x \end{pmatrix} [p_0(\theta,n,\phi)]^x \left[1-p_0(\theta,n,\phi) \right]^{\nu-x} .
\end{equation}
Here, $p_0(\theta, n,\phi)$ is the probability of measuring a single probe in the state $\ket{\Psi}$. In a two-dimensional noisy system, this probability is
\begin{equation}
    p_0(\theta, n,\phi)=\frac{1}{2}+\frac{\alpha \beta^n}{2} \cos(n\theta+\phi),
\end{equation}
where $\alpha,\beta$ are constants that parameterise the level of noise in the system. $\alpha=\beta=1$ in noiseless systems. 

\begin{figure}
    \centering
    \includegraphics[width = 8.5cm]{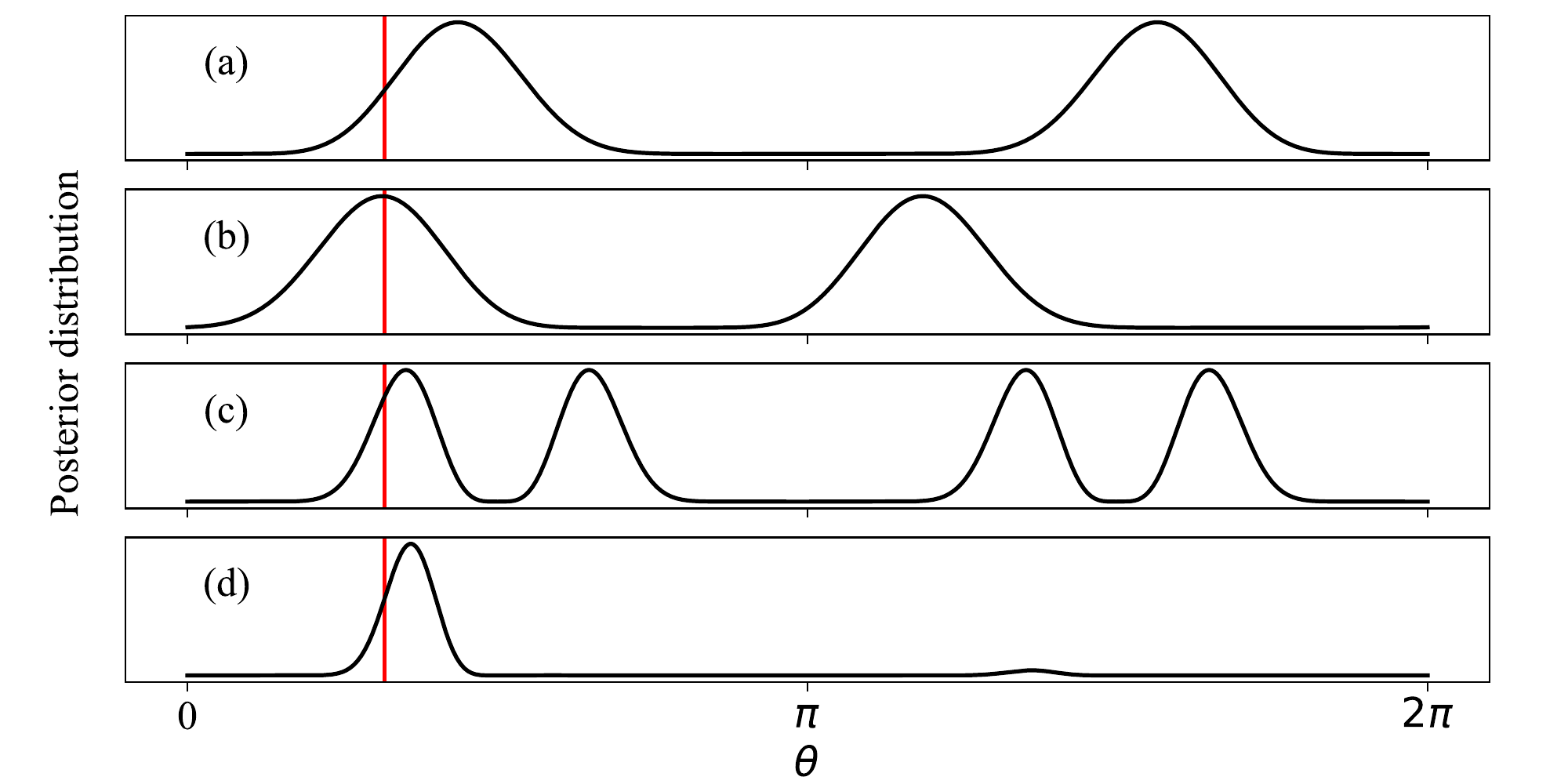}
    \caption{Posterior distribution when circuits \textbf{(a):} $(1,0)$, \textbf{(b):} $(1,\pi/4)$ and \textbf{(c):} $(2,0)$ are executed $10$ times with $\theta = 1$ [vertical line]. \textbf{(d):} By combining the results from (a)-(c), erroneous peaks can be reduced in size.}
    \label{fig:bayes}
\end{figure}

After observing $x$, Bayesian inference gives a posterior distribution for $\theta$ \cite{bayes_book}:
\begin{equation}
\label{eqn:bayes}
    p({\theta}|n,\phi,\nu,x)= \dfrac{p(x|n,\phi,\nu,{\theta})\pi({\theta})}{\int_0^{2\pi} p(x|n,\phi,\nu,{\theta})\pi({\theta}) d{\theta}},
\end{equation}
where $\pi(\theta)$ is any prior knowledge of $\theta$. This posterior can be used to produce an estimator of $\theta$, $\hat{\theta}$. Examples of such estimators include the minimum-mean-squared-error (MMSE) estimator, $\hat{\theta}_{\textrm{MMSE}}=\mathbb{E}\left[\theta \right]$, and the maximum-a-posteriori (MAP) estimator, $\hat{\theta}_{\textrm{MAP}}=\argmax_{{\theta}} p({\theta}|n,\phi,\nu,x)$. Alternatively, the posterior can be used to find the confidence that ${\theta}$ belongs to some interval $\Theta \subseteq [0,2\pi]$:
\begin{equation}
    \label{eqn:conf}
    \textrm{Pr} \left[ {\theta} \in {\Theta} \right] = \int_{{\Theta}} p({\theta}|{n,\phi},\nu,{x}) d {\theta} .
\end{equation} 
Some estimators can lack point-identification by taking multiple values \cite{point_id}. Lack of point-identification occurs during phase-estimation for $\hat{\theta}_{\textrm{MAP}}$ because $\cos^{-1}(n\theta)$ is multi-valued in the interval $[0,2\pi]$. This ambiguity is demonstrated in Figs. \ref{fig:bayes}(a)-(c): the posterior distribution has $2n$ symmetric peaks centred at either $\theta + \frac{2\pi l_1}{n}$ or $-\theta-\frac{2\phi}{n} + \frac{2\pi l_2}{n}$, where $l_1$ and $l_2$ are integers obeying $-\frac{n\theta}{2\pi} \leq l_1 \leq n - \frac{n\theta}{2\pi}$ and $\frac{n\theta}{2\pi} + \frac{\phi}{\pi} \leq l_2 \leq n + \frac{n\theta}{2\pi} + \frac{\phi}{\pi}$, respectively.

Point-identifying $\hat{\theta}_{\textrm{MAP}}$ (i.e. selecting the correct peak) can be achieved by executing several circuits, $(\boldsymbol{n},\boldsymbol{\phi}) \equiv \left\{(n_1,\phi_1),\dots,(n_m,\phi_m) \right\}$. Each circuit is executed a number $\boldsymbol{\nu}=(\nu_1,\dots,\nu_m)$ times, respectively, resulting in observations $\boldsymbol{x}=(x_1,\dots,x_m)$. This new measurement strategy applies $\hat{U}(\theta)$ a total $\sum_{j=1}^m n_j \nu_j$ times. Combining observations provides the posterior distribution
\begin{equation}
\label{eqn:new_bayes}
    p({\theta}|\boldsymbol{n},\boldsymbol{\phi},\boldsymbol{\nu},\boldsymbol{x})= \dfrac{ \pi({\theta}) \prod_{i=1}^{m} p(x_i|n_i,\phi_i,\nu_i,{\theta}) }{\int_0^{2\pi} \pi({\theta}) \prod_{i=1}^{m} p(x_i|n_i,\phi_i,\nu_i,{\theta}) d\theta} .
\end{equation}
Careful selection of $\boldsymbol{n}, \boldsymbol{\phi}$ and $\boldsymbol{\nu}$ can ensure that $\hat{\theta}_{\textrm{MAP}}$ is unique. This is demonstrated in Fig. \ref{fig:bayes}(d), which combines the observations of Figs. \ref{fig:bayes}(a)-(c).

\textit{Performance of estimators.---}The performance of an estimator is judged by an expected error. Other works often focus on the statistical variance of the individual peaks of the posterior distribution, $\sigma^2(\theta)$. $\sigma^2(\theta)$ can be estimated in the limit of many circuit executions, $\nu \to \infty$: If the single circuit $(n,\phi)$ is executed $\nu$ times, $p(x|n,\phi,\nu,\theta)$ tends to a normal distribution centred on $\nu p_0$ with variance $\nu p_0 ( 1 - p_0 )$ \cite{FI_tutorial}.
The posterior distribution is then the sum of $2n$ normal distributions, each with
\begin{equation}
\label{eqn:error}
{\sigma}^2(\theta) = \frac{1 - {\alpha^2\beta^{2n}} \cos^2(n\theta+\phi)}{\alpha^2\beta^{2n} \sin^2(n\theta+\phi)n^2 \nu},
\end{equation} 
estimated through linear error propagation.

\change{In the absence of noise, $\sigma^2(\theta) = \frac{1}{n^2 \nu}$ independent of $\theta$ and $\phi$. Restricting the number of times $\hat{U}(\theta)$ can be applied to $N_{\textrm{tot}}$ times, the circuit $(n,\phi)$ can be executed at most $\nu = \left\lfloor N_{\textrm{tot}}/n \right\rfloor$ times. Taking $n \propto N_{\textrm{tot}}$ in the limit $N_{\textrm{tot}} \to \infty$ results in a variance that scales according to the Heisenberg limit: $\sigma^2(\theta) = \left( 1/{N^2_{\textrm{tot}}}\right)$ \cite{HL}. Naturally, such infinitely long circuits are unrealistic. In real devices, circuit depth is restricted by some limiting value $n_{\textrm{lim}}$. Executing the circuit $(n_{\textrm{lim}},\phi)$ results in the  standard quantum limit: $\sigma^2(\theta)=O\left( 1/{N_{\textrm{tot}}}\right)$. 

When noise is present, $\sigma^2(\theta)$ is minimized when the circuit
\begin{equation}
\label{eqn:opt_circ}
    \left( n_{\textrm{opt}}, \phi_{\textrm{opt}}\right) = \left(- \dfrac{1}{2 \ln \beta}, \dfrac{\pi}{2} -n_{\textrm{opt}}\theta \right) ,
\end{equation}
is executed: $\min_{n,\phi} {\sigma}^2(\theta) = - \frac{2e\ln \beta}{\alpha^2 N_{\textrm{tot}}}$. This variance scales with the standard quantum limit, $O\left( 1/{N_{\textrm{tot}}}\right)$. However, the pre-factor is reduced by $-2\beta^2 e \ln \beta$ compared to the purely classical scaling. This reduction constitutes a quantum advantage. A complication is that the value of $\phi_{\textrm{opt}}$ depends on $\theta$ so is unknown before sampling begins. Therefore, one has to rely on an adaptive scheme to find $\phi_{\textrm{opt}}$.
}

The aforementioned variance is often quoted as the figure of merit of phase-estimation algorithms \cite{Cramer}. However, this is only appropriate if the correct peak (out of the $2n$ peaks) is picked out with certainty. This selection is typically achieved by sampling shallower circuits alongside the optimal circuit. Doing this requires the applications of $\hat{U}(\theta)$, and thus prevents $\sigma^2(\theta)$ from reaching its minimum value for a given amount of resources. 

We argue that the variance is a poor figure of merit, as it ignores errors arising from smaller, secondary peaks in the posterior distribution, such as the one in Fig. \ref{fig:bayes}(d). Instead, we argue in favour of using the expected value of a loss function $L(\hat{{\theta}},{\theta})$ as the figure of merit of an estimator's performance:
\begin{equation}
\label{eqn:estimated_loss}
    {\mathcal{L}}(\hat{\theta})=\mathbb{E} \left[ L(\hat{\theta},\theta) \right]=\int_{0}^{2\pi} p({\theta}|\boldsymbol{n},\boldsymbol{\phi},\boldsymbol{\nu},\boldsymbol{x}) L(\hat{{\theta}},{\theta}) d{\theta}.
\end{equation}
Examples of loss functions include the absolute error, $|\hat{{\theta}}-{\theta}|$, and the squared error, $(\hat{{\theta}}-{\theta})^2$.  To our knowledge, no iterative phase-estimation algorithm has observed Heisenberg scaling of the mean absolute error or the root mean squared error. Below, we present an algorithm which achieves these scalings.

\textit{Adaptive phase-estimation algorithm.---}We now present a phase-estimation algorithm which achieves Heisenberg scaling of the mean absolute error (MAE) and the mean squared error (MSE). It does so by iterating through  circuits of doubling depths, similar to previous algorithms \cite{RG, Me, iterative, iterative2, iterative3}. Our algorithm differs from these in its adaptive nature, selecting which circuit to execute next depending on the outcomes of previous measurements and the number of resources left. The resource considered is the total number of times $\hat{U}(\theta)$ is applied throughout the algorithm, which we fix to $N_{\textrm{tot}}$ before the algorithm commences. After a circuit is sampled, the posterior distribution $p(\theta|\boldsymbol{n},\boldsymbol{\phi},\boldsymbol{\nu},\boldsymbol{x})$ [Eq. \eqref{eqn:new_bayes}] is recalculated, and a new estimate $\hat{\theta}$ is produced. These  are used for calculating confidence levels [Eq. \eqref{eqn:conf}] and for predicting future loss. 

The iteration is structured as follows: At the $i^{\textrm{th}}$ step, the circuit $(n_{i},\phi_i)$ is sampled until $\mathrm{Pr} \left[ \theta \in \Theta_i \right] \geq 1 - \epsilon_i $. We set the circuit depth  as $n_i = \min(2^{i-1},n_{\textrm{opt}},n_{\textrm{lim}})$, to ensure that, within the optimal circuit depth, the next circuit is twice as deep as the previous one. The phase shift is set to $\phi_i = \frac{\pi}{2} - n_i \theta_c$, where $\theta_c$ is the centre of the interval $\Theta_i = \left[  {\theta}_{c} - \frac{\pi}{2n_{i+1}} , {\theta}_{c} + \frac{\pi}{2n_{i+1}} \right]$: 
\begin{equation}
    {\theta}_{\textrm{c}} = \begin{cases}
    \theta_{\textrm{min}} + \frac{\pi}{2n_{i+1}} & \hat{\theta} < \theta_{\textrm{min}} + \frac{\pi}{2n_{i+1}} \ \textbf{and} \ i \neq 1\\
    \theta_{\textrm{min}} - \frac{\pi}{2n_{i+1}} & \hat{\theta} > \theta_{\textrm{max}} - \frac{\pi}{2n_{i+1}} \ \textbf{and} \ i \neq 1 \\
    \hat{\theta} & \textrm{otherwise}.
    \end{cases}
\end{equation}
Here, $\Theta_{i-1} = \left[\theta_{\textrm{min}},\theta_{\textrm{max}} \right]$ is the interval in the $(i-1)^{\textrm{th}}$ step. We choose $\Theta_i$ in this this way to ensure that: (1) all estimates from the successive circuit $(n_{i+1},\phi_{i+1})$ are point-identified because $\cos^{-1}\left( n_{i+1} \theta +\phi_{i+1} \right)$ is single valued in $\Theta_i$; and (2) $\Theta_i$ never extends beyond the limits of $\Theta_{i-1}$, unless $i=1$ where the discontinuity at $\theta=2\pi$ can be traversed.
The values of $\epsilon_i$ must follow $\epsilon_i < \epsilon_{j}$ for all $i<j$ such that  $\textrm{Pr}\left[\theta \in \Theta_j \right] \leq \textrm{Pr}\left[\theta \in \Theta_i \right]$.
In our simulations, presented below, we  use $\epsilon_i = \left( \frac{n_i}{N_{\textrm{tot}}} \right)^3$. In Appendix \ref{append_scale}, we prove that this choice leads to Heisenberg scaling of the MAE and MSE.  

After enough probes have been measured such that $\mathrm{Pr} \left[ \theta \in \Theta_i \right] \geq 1 - \epsilon_i $, we use the current posterior distribution to determine the next step. Our algorithm calculates the predicted loss of  using all of the remaining resources to execute either the circuit $(n_i,\phi_i)$ or $(n_{i+1},\phi_{i+1})$. The circuit with the smallest predicted loss is then executed.  If $(n_{i+1},\phi_{i+1})$ is [is not] selected, the iteration proceeds with $i \rightarrow i+1$ [iteration concludes with the circuit $(n_i,\phi_i)$ executed until all remaining resources are used up]. Losses are predicted by estimating expected outcomes. Consider an arbitrary point in the algorithm where the posterior distribution is $p(\theta|\boldsymbol{n},\boldsymbol{\phi},\boldsymbol{\nu},\boldsymbol{x})$ and $\hat{U}(\theta)$ can still be applied a number $N_{\textrm{left}}$ times. Any circuit $(n,\phi)$ can be executed up to $\nu_{(n,\phi)} = \left\lfloor \frac{N_{\textrm{left}}}{n} \right\rfloor$ times. The expected number of measurements corresponding to the outcome $\ket{\Psi}$ is
\begin{equation}
\begin{split}
    \hat{x}_{(n,\phi)} = \int_{0}^{2\pi} \nu_{n} p_0(\theta, n, \phi) p(\theta|\boldsymbol{n},\boldsymbol{\phi},\boldsymbol{\nu},\boldsymbol{x}) d\theta.
\end{split}
\end{equation}
$\hat{x}_{(n,\phi)}$ is used to generate a new posterior distribution from which the expected loss $\hat{\mathcal{L}}_{(n,\phi)}$ is calculated. 

\begin{figure}
    \centering
    \includegraphics[width=8cm]{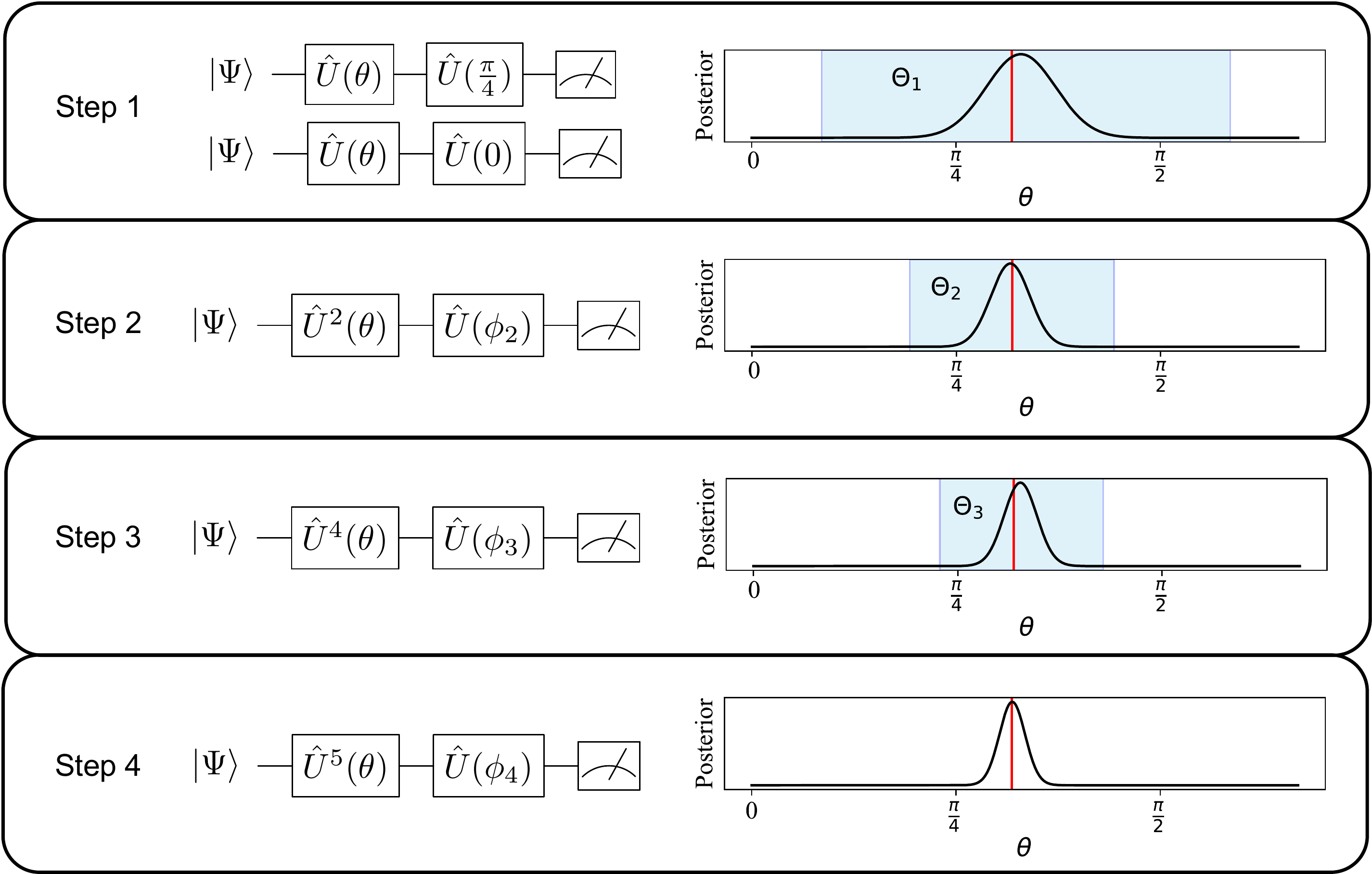}
    \caption{The circuits executed at each iteration step of our algorithm and the resultant posterior distribution for initial conditions $\theta = 1$ [vertical line], $N_{\textrm{tot}}=300$ and $\beta = 0.9$. The shaded regions represent the different intervals $\Theta_i$. 
    }
    \label{fig:algorithm_circuits}
\end{figure}

Our algorithm can be summarised as follows:
\begin{enumerate}

    \item Execute the two circuits $(1,0)$ and $(1,\pi/4)$ until $\textrm{Pr} \left[\theta \in \Theta_1 \right] \geq 1-\epsilon_1$.

    \item Calculate the expected losses $\hat{\mathcal{L}}_{(1,\phi_1)}$ and $\hat{\mathcal{L}}_{(2,\phi_2)}$.

        \item \textbf{If} $\hat{\mathcal{L}}_{(1,\phi_1)}<\hat{\mathcal{L}}_{(2,\phi_2)}$, skip to \ref{step:last}. 

    \item \textbf{Else}, set $i=2$, and the current circuit to $(n_i,\phi_i)$.

    \item While $N_{\textrm{left}}>n_i$, \textbf{do}:
    \begin{enumerate}
        \item Execute the current circuit until $\textrm{Pr} \left[\theta \in \Theta_1 \right] \leq 1-\epsilon_i$. 
        \label{step:loop}
        \item Calculate $\hat{\mathcal{L}}_{(n_i,\phi_i)}$ and $\hat{\mathcal{L}}_{(n_{i+1},\phi_{i+1})}$.
        \item \textbf{If} $\hat{\mathcal{L}}_{(n_{i+1},\phi_{i+1})} < \hat{\mathcal{L}}_{(n_{i},\phi_{i})}$, set the current circuit to $(n_{i+1},\phi_{i+1})$ and $i \leftarrow i + 1$. Return to \ref{step:loop}.
        \item \textbf{Else}, update $\phi_i \rightarrow \frac{\pi}{2}-n_i \hat{\theta}$ after each circuit execution and  sample the circuit $(n_i,\phi_i)$ a number $\left\lfloor \frac{N_{\textrm{left}}}{n_i} \right\rfloor$ times.
        \end{enumerate}
    
    \item Execute circuit $(1,\phi_1)$ a number $N_{\textrm{left}}$ times to use up any remaining resources. \label{step:last}

\end{enumerate}
The first step involves execution of two circuits, $(1,0)$ and $(1,\pi/4)$, to reduce the size of the additional peak $2\pi - \theta$. Figure \ref{fig:algorithm_circuits} pictorially shows the circuits sampled, and the resultant posterior distribution, at different stages of the algorithm.

Our adaptive algorithm does not suffer from the same shortfalls as other non-adaptive phase-estimation algorithms: (1) Other algorithms use multiple different fixed intervals instead of our adaptively changing single interval $\Theta_i$. In these previous algorithms, if $\theta$ is close to a  boundary between intervals, a large number of circuit executions is needed to place $\theta$, with high confidence, within one of the intervals. Indeed if $\theta$ lies on the boundary, circuits must be executed infinite number of times \cite{Me}. (2)  Other algorithms use pre-decided constant numbers of circuit executions. If $\theta$ is far from the interval boundary, this results in building over-confidence in that $\theta \in \Theta_i$. Consequently, these algorithms do not use the resources optimally. (3) Previous algorithms lack a cut-off depth, such that circuits with depths exceeding $n_{\textrm{opt}}$ are sampled. (4) Previous algorithms do not take into account hardware limitations on circuit depths.

\begin{figure}
    \centering
    \includegraphics[width=8cm]{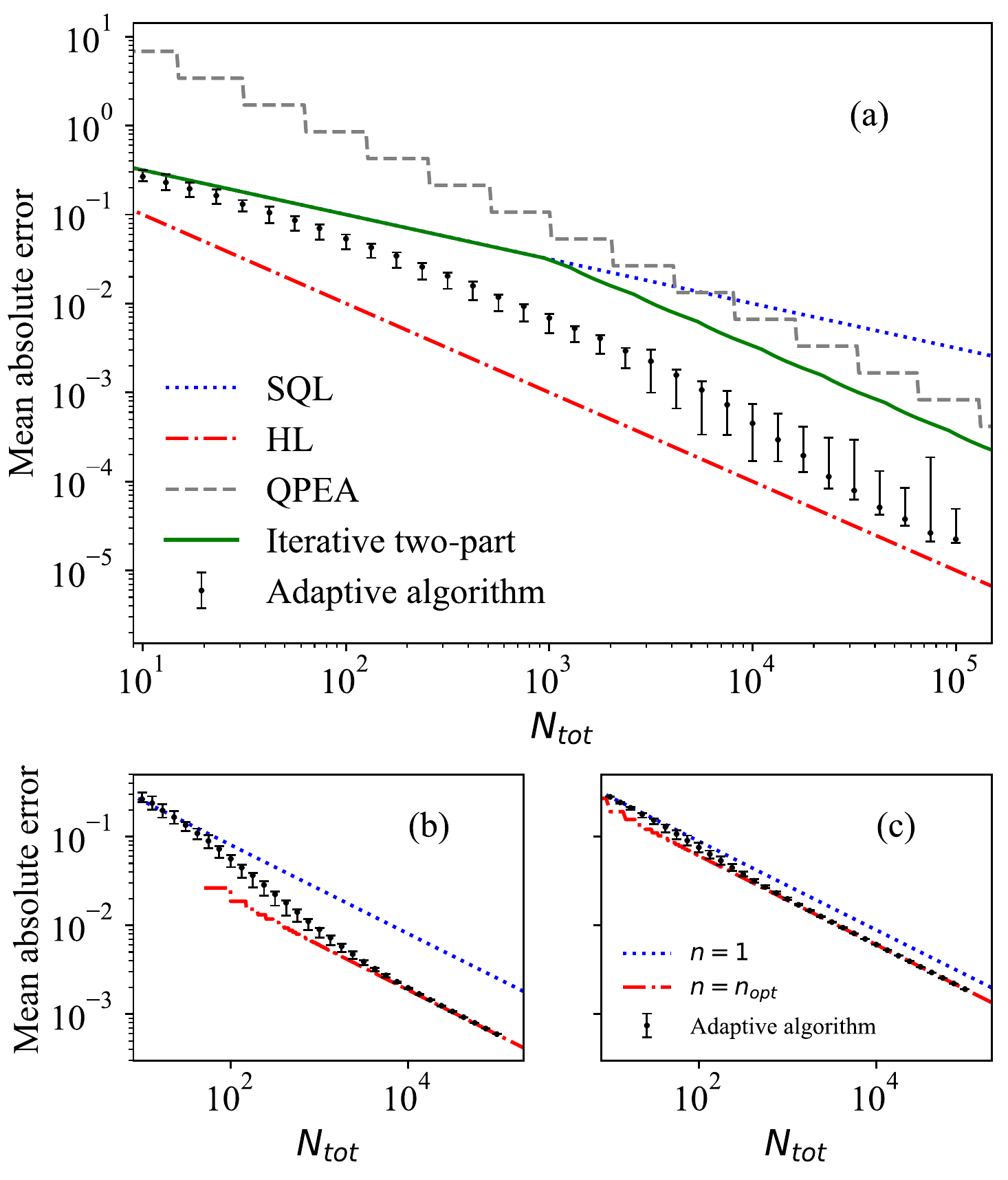}
    \caption{Simulated values of the mean absolute error against total number of unitary applications for different phase-estimation algorithms for $\alpha = 1$ and: (a) $\beta = 1$; (b) $\beta = 0.99$; (c) $\beta = 0.9$. Plots (b) and (c) use the same legend. Error bars are plotted between the minimum and maximum mean absolute error achieved after 100 samples.}
    \label{fig:error_plot}
\end{figure}

\textit{Simulations.---}We benchmark the performance of our Bayesian algorithm through numerical simulations. For a given $N_{\textrm{tot}}$, the algorithm is evaluated for $100$ evenly spaced values of $\theta$ in the range $[0,2\pi]$ and $\mathcal{L}(\hat{\theta})$ is calculated. Figure \ref{fig:error_plot} is generated using the mean absolute error (MAE) as the expected loss for $\hat{\theta}_{\textrm{MAP}}$ and assuming a uniform prior. We generate plots of the MAE against $N_{\textrm{tot}}$  for different values of $\beta$. 

Figure \ref{fig:error_plot}(a) plots the MAE performance when noise does not scale with circuit depth ($\beta = 1$) and $n_{\textrm{lim}}$ exceeds $10^6$. We also plot data from the standard quantum limit (SQL), the Heisenberg limit (HL), the quantum-phase-estimation algorithm (QPEA) \cite{QPEA} and the non-adaptive iterative two-part protocol \cite{Me}. Our adaptive algorithm has a sub-SQL MAEs for a $N_{\textrm{tot}}$ greater than roughly $20$. The iterative two-part and QPEA have sub-SQL MAEs for $N_{\textrm{tot}}$ greater than $1,000$ and $4,000$, respectively. This demonstrates that our Bayesian algorithm achieves a quantum advantage with fewer resource used than the other two popular algorithms. At the point where the QPEA demonstrates quantum advantage, our algorithm achieves an MAE that, at worst, is ten times smaller.

In Figures \ref{fig:error_plot}(b)-(c), we plot the MAE performance when noise does scale with circuit depth: $\beta < 1$. We also plot $\sigma^2(\theta)$ for the circuit $(1,\phi_1)$ [blue dots] and the circuit $(n_{\textrm{opt}},\phi_{\textrm{opt}})$ [red dot-dash]. Our Bayesian algorithm tends to to the optimal limit when $N_{\textrm{tot}}$ is large. We also observe a reduced spread of the MAE achieved for larger $N_{\textrm{tot}}$, indicating the algorithms ability to find the optimal circuit.

\newstuff{


}

\textit{Conclusions.---}We have presented an adaptive algorithm for quantum phase estimation based on Bayesian inference. Our algorithm samples a series of circuits with parameters determined, iteratively, by the outcomes of previous measurements. The algorithm operates to minimize the predicted expected loss for a given  amount of resources. Our algorithm achieves the Heisenberg scaling of both the mean absolute error and the mean squared error in the noiseless setting, an achievement previously only found in the QPEA, which requires inter-probe entanglement.  Furthermore, in the presence of noise, our algorithm's performance tends to the optimal error limit imposed by the statistical variance. 

The authors thank N. Mertig and W. Salmon for enlightening discussions. 
This work was supported by Hitachi, Lars Hierta's Memorial Foundation, and Girton College, Cambridge.

\bibliography{paper}

\appendix

\section{Asymptotic Scaling}
\label{append_scale}

We now derive the asymptotic complexity of our adaptive algorithm, assuming a uniform prior. This limit ensures circuits $(\boldsymbol{n},\boldsymbol{\phi})$ are executed as $\boldsymbol{\nu} \to \infty$. 

After the $i^{\textrm{th}}$ iteration step, the circuits $(n_1,\phi_1),\dots, (n_m,\phi_m)$ have been executed $\nu_1,\dots,\nu_m$ times, and the current posterior distribution is 
\begin{equation}
     N \left( \theta, \Sigma_i^2 \right) = N \left( \theta, \dfrac{1}{\sum_{j=1}^{i} \sigma_j^{-2}(\theta) } \right)
\end{equation}
plus some minor peaks, where $\sigma_j^2(\theta)$ is the variance for the circuit $(n_j,\phi_j)$ [Eq. \eqref{eqn:error}]. $\nu_i$ has a maximum value in order to achieve $\textrm{Pr} \left[\theta \in \Theta_i \right] \geq 1 - \epsilon_i$ which can be derived by the Chernoff-Cramér bound \cite{nielsen}:
\begin{equation}
\label{eqn:chernoff}
    \textrm{Pr} \left[ |\hat{\theta}-\theta| \geq \delta_i \right] = \epsilon_i \leq 2 \exp{ \left[ - {\delta_i^2} / \Sigma_i^2 \right]},
\end{equation}
where $\delta_i=\frac{\pi}{2n_{i+1}}$, half the length of $\Theta_i$. Rearranging, $\Sigma^{-2}_i \geq \frac{8n^2_{i+1}}{\pi^2} \ln \left( \frac{2}{\epsilon_i} \right)$, with worst case performance occurs for equality (assumed from now on). By noting $\Sigma^{-2}_i = \Sigma^{-2}_{i-1} + \sigma^{-2}_{i}$, we find:
\begin{equation}
    \sigma_i^2(\theta) = \frac{8}{\pi^2} n^2_{i+1} \left[ \ln \left( \frac{2}{\epsilon_i} \right) - \frac{1}{4} \ln \left( \frac{2}{\epsilon_{i-1}} \right) \right].
\end{equation} 
Equating this to Eq. \eqref{eqn:error},
\begin{equation}
    \nu_i = \frac{32}{\pi^2 \alpha^2 \beta^{2n_i}} \left[ \ln \left( \frac{2}{\epsilon_i} \right) - \frac{1}{4} \ln \left( \frac{2}{\epsilon_{i-1}} \right) \right]
\end{equation}
for $i < m$.

We consider profiles of $\epsilon_i = \epsilon \left(\frac{n_i}{n_m} \right)^p$, where $\epsilon$ is a positive constant. This form of polynomial ensures $\epsilon_j \leq \epsilon_k$ for all $j < k$, and results in
\begin{equation}
    \ln \epsilon_i = p \ln n_i - p \ln n_m = (i - 1) p \ln 2 - p \ln n_m
\end{equation}
for $i<m$. The total number of times $\hat{U}(\theta)$ is applied is fixed to $N_{\textrm{tot}}$. Therefore,
\begin{equation}
\begin{split}
    N_{\textrm{tot}} &= \sum_{j=1}^m n_j \nu_j \\ 
    &= n_m \nu_m + \sum_{j=1}^{m-1}  \frac{2^{i+4}}{\pi^2 \alpha^2\beta^{2^i}} \left[ \frac{3}{4} \ln2 - \ln \epsilon_i + \frac{1}{4} \ln \epsilon_{i-1} \right] 
\end{split}
\end{equation}
When $\beta < 1$, or $n_{\textrm{lim}}$ is finite, $m$ is restricted to some finite value and, hence, the sum is a constant independent of $m$. If, however, $\beta=1$ and $n_{\textrm{lim}} \to \infty$, and $n_m=2^m$. The above sum scales as $O(2^m)$ for large $m$. Therefore,
\begin{equation}
\label{eqn:n_t}
    n_m \nu_m = N_{\textrm{tot}} - \begin{cases}
    c_1 & \beta < 1 \ \textrm{or} \ n_{\textrm{lim}} < \infty \\
    c_2 2^m & \textrm{otherwise},
    \end{cases}
\end{equation}
where $c_1$ and $c_2$ are constants. Equation \eqref{eqn:n_t} can be used to find how $\Sigma^2_m$ scales with large $m$:
\begin{equation}
\label{eqn:Sigma}
\begin{split}
    \Sigma^{-2}_m &= \Sigma_{m-1}^{-2} + \sigma^{-2}_m\\
    &= \frac{8n^2_{m}}{\pi^2} \ln \left( \frac{2}{\epsilon_{m-1}} \right) + \alpha^2 \beta^{2n_m} n_m^2 \nu_m \\
    &= c_3 n_m N_{\textrm{tot}} - c_4 n_m^2 
\end{split}
\end{equation}
where $c_3$ and $c_4$ are constants.

After the algorithm concludes, the expected loss [Eq. \eqref{eqn:estimated_loss}] can be bounded by splitting the integral over $[0,2\pi]$ into the sum of integrals over the different intervals $\Theta_i$: 
\begin{equation}
\label{eqn:loss_a}
\begin{split}
    \mathcal{L}(\theta) &= \sum_{i}^{m-1} \int_{\bar{\Theta}_i} p(\theta|\boldsymbol{n}, \boldsymbol{\phi}, \boldsymbol{\nu}, \boldsymbol{x}) L(\hat{\theta},\theta) d\theta \\
    & \quad + \int_{\Theta_m} p(\theta|\boldsymbol{n}, \boldsymbol{\phi}, \boldsymbol{\nu}, \boldsymbol{x}) L(\hat{\theta},\theta) d\theta,
\end{split}
\end{equation}
where $\bar{\Theta}_i = \Theta_{i-1} - \Theta_i$. By definition,
\begin{equation}
\begin{split}
    \int_{\bar{\Theta}_i} p(\theta|\boldsymbol{n}, \boldsymbol{\phi}, \boldsymbol{\nu}, \boldsymbol{x}) d\theta &= \int_{{\Theta}_{i-1}} p(\theta|\boldsymbol{n}, \boldsymbol{\phi}, \boldsymbol{\nu}, \boldsymbol{x}) d\theta \\
    & \quad - \int_{{\Theta}_i} p(\theta|\boldsymbol{n}, \boldsymbol{\phi}, \boldsymbol{\nu}, \boldsymbol{x}) d\theta \\
    &= \epsilon_i - \epsilon_{i-1}.
\end{split}
\end{equation}
Therefore, each integral in Eq. \eqref{eqn:loss_a} over the interval $\bar{\Theta}_i$ is upper bounded by $\epsilon_i - \epsilon_{i-1}$ multiplied by the maximum of $L(\hat{\theta},\theta)$ over $\hat{\Theta}_i$:
\begin{equation}
    \max_{\theta \in \bar{\Theta}_i} L_{\textrm{MAE}}(\hat{\theta},\theta) = |\Theta_i| \ \ \max_{\theta \in \bar{\Theta}_i} L_{\textrm{MSE}}(\hat{\theta},\theta) = |\Theta_i|^2.
\end{equation}

Overall, we can bound the mean absolute error (MAE) and mean squared error (MSE) as follows:
\begin{equation}
\label{eqn:loss_bounds}
\begin{split}
    \mathcal{L}_{\textrm{MAE}}(\hat{\theta}) &\leq 2\pi \epsilon_1 + \sum_{i=2}^{m-1} (\epsilon_{i}-\epsilon_{i-1}) \frac{\pi}{n_i} + \sqrt{\frac{2 \Sigma^2_m}{\pi}} \\
    &= \frac{3\pi \epsilon_1}{2} + \sum_{i=2}^{m-1} \epsilon_{i} \frac{\pi}{2n_i} + \sqrt{\frac{2 \Sigma^2_m}{\pi}} \\
    \mathcal{L}_{\textrm{MSE}}(\hat{\theta}) &\leq 4 \pi^2 \epsilon_1 + \sum_{i=2}^{m-1} (\epsilon_{i}-\epsilon_{i-1}) \frac{\pi^2}{n_i^2} + \Sigma_m^2 \\
    &= \frac{15\pi^2 \epsilon_1}{4} + \sum_{i=2}^{m-1} \epsilon_{i} \frac{3\pi^2}{4n_i^2} + \Sigma_m^2.
\end{split}
\end{equation}
The additional terms linear in $\epsilon_i$ can be arises from incorrect point-identification. For a constant $\epsilon_i$ over all circuits, the $\epsilon_1$ term leads to a constant, such that the loss scales as $O(N^0_{\textrm{tot}})$. Using Eq. \eqref{eqn:Sigma}:
\begin{equation}
\begin{split}
\mathcal{L}_{\textrm{MAE}}(\hat{\theta}) &\leq \sqrt{\frac{2}{c_3 \pi n_m N_{\textrm{tot}} -  c_{4} \pi n_m^2}} \\ & \quad \quad + c_5 \epsilon \begin{cases} 
    2^{-pm} & p < 1 \\
    m 2^{-m} & p = 1 \\
    2^{-m} & p > 1.
    \end{cases} \\
\mathcal{L}_{\textrm{MSE}}(\hat{\theta}) &\leq \frac{1}{{c_3 n_m N_{\textrm{tot}} - c_{4} n_m^2}} \\ & \quad \quad + c_6 \epsilon \begin{cases} 
    2^{-pm} & p < 2 \\
    m 2^{-2m} & p = 2 \\
    2^{-2m} & p > 2.
    \end{cases}
\end{split}
\end{equation}
Choosing $p = 3$, the MAE scaling becomes
\begin{equation}
\begin{split}
\mathcal{L}_{\textrm{MAE}}(\hat{\theta}) = \begin{cases}
    O \left(\frac{1}{\sqrt{ N_{\textrm{tot}}} } \right) & \beta < 1 \ \textrm{or} \ n_{\textrm{lim}} < \infty \\
    O \left(\frac{1}{N_{\textrm{tot}}} \right) & \textrm{otherwise} 
\end{cases}
\end{split}
\end{equation}
and the MSE scaling becomes
\begin{equation}
\begin{split}
\mathcal{L}_{\textrm{MAE}}(\hat{\theta}) = \begin{cases}
    O \left(\frac{1}{{ N_{\textrm{tot}}} } \right) & \beta < 1 \ \textrm{or} \ n_{\textrm{lim}} < \infty \\
    O \left(\frac{1}{N^2_{\textrm{tot}}} \right) & \textrm{otherwise}
\end{cases}
\end{split}
\end{equation}
if we set $N_{\textrm{tot}}=2^m$. 

Therefore, the choice
\begin{equation}
    \epsilon_i \propto \left( \frac{n_i}{N_{\textrm{tot}}} \right)^3
\end{equation}
provides Heisenberg scaling of the MAE and MSE in noiseless scenarios.

\end{document}